\documentclass[10pt,twocolumn]{revtex4}
\usepackage{graphicx}
\usepackage{psfig}
\newcommand{\be}{\begin{equation}}
\newcommand{\bea}{\begin{eqnarray}}
\newcommand{\ee}{\end{equation}}
\newcommand{\eea}{\end{eqnarray}}
\newcommand{\eps}{\epsilon}

\begin{document}
\title{Dynamics of a many-particle Landau-Zener model: inverse sweep}
\author{A.P. Itin$^{1,2}$ and  P. T\"{o}rm\"{a}$^1$ }
\address{
$^{1}$Department of Applied Physics, Helsinki University of
Technology, P.O. Box 5100, 02015, Finland \\ $^{2}$Space Research
Institute, RAS, Profsoyuznaya str. 84/32, 117997 Moscow, Russia }
\begin{abstract}
We consider dynamics of a slowly time-dependent Dicke model, which
represents a many-body generalization of the Landau-Zener model.
In particular, the model describes narrow Feshbach resonance
passage in an ultracold gas of Fermi atoms. Adiabaticity is
destroyed when a parameter crosses a critical value, even at very
slow sweeping rates of a parameter. The dynamics crucially depends
on direction of the sweep. We apply our recent analysis [A.P.
Itin, P. Torma, arXiv:0901.4778v1] to the "inverse" sweep through
the resonance, corresponding (in a context of Feshbach resonance
passage) to dissociation of molecules. On a level of the
mean-field approximation, the dynamics is equivalent to a
molecular condensate formation from Bose atoms within a two-mode
model. Mapping the system to a Painlev\'{e} equation allows us to
calculate deviation from adiabaticity at very slow sweeps
analytically.
\end{abstract}

\maketitle

Destruction of adiabaticity is conceptually important issue in
both quantum and classical mechanics. Nonadiabatic dynamics of many quantum single-particle systems can
be effectively described in terms of the  exactly solvable
Landau-Zener model \cite{LZ} (LZM), where a probability of
transition from an initially occupied instantaneous ground state
level to an excited one is exponentially small in a sweeping rate
parameter. Numerous studies have been devoted to such type of
models \cite{q1}.

Many-particle quantum systems are intriguing objects that can
exhibit nonlinear behavior and in some sense link quantum and
classical worlds together. Interest to such systems has grown
enormously due to progress in ultracold quantum gases experiments
\cite{exp2}. Destruction of adiabaticity in semiclassical models
of many-particle systems under slow driving has been intensively
investigated recently
\cite{Zobay,Javanainen,Pazy,Gurarie,Tokyo,PhysD} using various
methods. It has been found that exponential LZM-type behavior for
transition probabilities is often replaced with power-laws of
sweeping rates. Theoretical methods used for derivation of these
power-laws often involve a number of
approximations those accuracy is difficult to control.

We considered recently \cite{ITorma} a time-dependent Dicke model
(see Eq. \ref{Polkovnikov} below) using concepts of classical
adiabatic invariants. Technically, a method which naturally links
classical mechanics and quantum dynamics of many-particle systems
is a so-called Truncated Wigner Approximation (TWA)
\cite{TWA,TWA2,Gurarie}. Briefly, within this approximation one
considers an ensemble of classical trajectories whose initial
conditions are distributed in phase space according to a Wigner
transform of an initial quantum state. One therefore is left with
the problem of nonadiabaticity of individual classical solutions.
In classical mechanics, it is known that adiabaticity is destroyed
when a phase point crosses or comes close to a separatrix of an
unperturbed system \cite{AKN}. There is a method for calculating
changes in approximate adiabatic invariants: separatrix crossing
theory \cite{Cary,N86,AKN}. The method has a drawback: usually it is not
applicable in case a separatrix crossing happens close to or at
the time of bifurcation, i.e. in case of very small initial action
of a classical trajectory. In \cite{ITorma}, we managed to
overcome this drawback by mapping the system close to a
bifurcation to a Painlev\'{e} equation, those asymptotics are known.
A formula for change in the action we derived there allowed to
describe mean values of number of bosons created at the resonance
passage, and its distributions. In this Brief Report we consider
the "inverse" sweep through the resonance in the same model. We
also discuss other physical realizations of the semiclassical
model, in particular association of Bose atoms in diatomic
molecules within a simple two-model model \cite{Javanainen}.

Let us consider the time-dependent Dicke model written in the
following form \cite{Barankov, Gurarie}:
\begin{equation} \hat{H} = - \Delta(t) b^\dagger b + \Delta(t)
S^z + \frac{g}{\sqrt{N}}(b^{\dagger} S^- + b S^+), \label{Dicke}
\end{equation} where $\frac{g}{\sqrt{N}}$ is the coupling strength,
$S^{\pm}=S_x \pm i S_y $ are spin operators, $b^\dagger$ and $b$
are creation and destruction operators of a bosonic mode,
$\Delta(t)= \pm\eps t$ is detuning, $\epsilon$ is the sweeping
rate of the bosonic mode energy. $\Delta(t)=+\eps t$ for a
"forward" sweep through resonance, and $-\eps t$ for a "backward", or "inverse",
sweep. The spin value $S$ is macroscopically large $S=N/2 \gg 1$.
With $N=2$, one recovers the standard Landau-Zener model. We set
$g=1$ for convenience in this paper.

For a "forward" sweep through the resonance considered in
\cite{ITorma}, we started in the distant past with some small
initial number of bosons ${N_b(t)|}_{-\infty} \equiv \langle
b^\dagger b \rangle (t)|_{-\infty} =N_- \equiv {\bf{n_-}} N$, and
calculated the final number of bosons $N_b(t)|_{+\infty} \equiv
{\bf{n_f}} N$ by averaging over an ensemble of classical
trajectories. For a backward sweep, only numerical results were
presented in Altland et al \cite{Gurarie}, consistent with the
earlier work of \cite{Javanainen}. Our method enables to explain
them analytically with high accuracy.
\begin{figure}
\includegraphics[width=78mm]{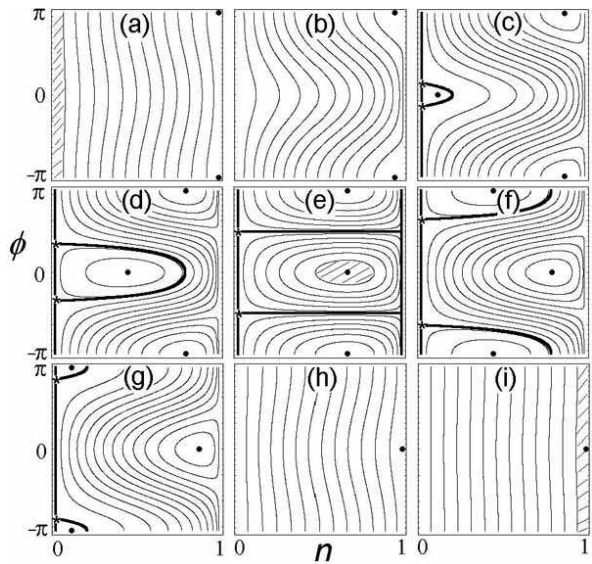}
 \caption{Phase portraits of the Hamiltonian (\ref{Polkovnikov}). From (a) to (f): $\gamma=-20,-4,-1.8,-0.9,0.,0.9,1.8,20,100$ correspondingly.
 Saddle points are denoted by asterisks, the bold trajectory is the separatrix. Shaded areas illustrate definitions of the
classical actions (see main text). }
\end{figure}
The classical limit of (\ref{Dicke}), obtained using the
$c-number$ formalism \cite{Gurarie},\cite{PhysD}, is
\begin{equation} H=-\gamma n - 2n \sqrt{1-n}\cos \phi, \quad
\gamma= \pm 2 \epsilon t, \label{Polkovnikov} \end{equation} where
$n$ corresponds to the number of bosons, and $\phi$ is the
canonically conjugated phase. For a forward sweep, $\gamma=  2
\epsilon t$, while for a backward sweep $\gamma= - 2 \epsilon t$.
Note that by means of a trivial change of variables
\cite{footnote}, the Hamiltonian becomes the same as analyzed in
\cite{PhysD} within the same context, and can be derived from
semiclassical Bloch equations of \cite{Pazy}. A mean-field model
of association of bosonic condensates into a molecular condensate
considered in \cite{Javanainen} is also equivalent to
(\ref{Polkovnikov}), with $n=1$ corresponding to "all-atom" mode
(i.e., on a mean-field level association of bosonic atoms is
analogous to dissociation of molecular condensate of Fermi atoms
\cite{footnote2}).

Let us  recall the classical phase space of the Hamiltonian
(\ref{Polkovnikov}) at fixed values of the parameter $\gamma$, as
explained in \cite{PhysD}. If $\gamma <-2$, there is only one
stable elliptic point on the phase portrait (at $\phi = \pi$ and
$n$ not far from $1$, see Fig.1a,b). At $\gamma = -2$, a
bifurcation takes place, and at $|\gamma| < 2$ phase portraits
look as shown in Figs.1c,d,e,f,g. There are two saddle points at
$n = 0, \cos\phi=-\gamma/2$, and a newborn elliptic point at $\phi
= 0$. The trajectory connecting these two saddles separates
rotations and oscillating motions and it is called the {\em
separatrix} of the frozen system \cite{PhysD}. Finally, at large
positive $\gamma$, again there is only one elliptic stationary
point at $\phi = 0$, and $n$ close to $1$. The classical action is
defined as in \cite{PhysD,Gurarie} and is shown graphically in
Figs 1a,e,i: shaded areas, divided by 2$\pi$. At $\gamma=-\infty$,
the action coincides with $n$: $I=n$. At $\gamma=+\infty$, we have
$I=1-n$. For a forward sweep, starting at $t=-\infty$ with certain
small $n=n_-$, in the adiabatic limit we get $n_+ = 1-n_-$ at $t =
+\infty$. Analogously, for a backward sweep, starting from a $n_-
\approx 1 $, in the adiabatic limit we get $n_+ = 1-n_-$ at $t =
+\infty$. Destruction of adiabaticity in both cases is mainly
caused by motion near the saddle points located at $n=0$, i.e. no
matter in what direction a sweep happens, region of the loss of
adiabaticity is the same.
\begin{figure}
\includegraphics[width=78mm]{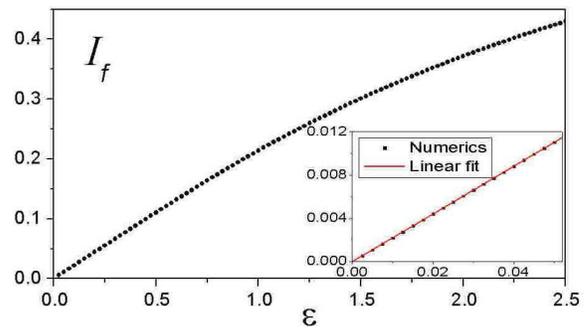}
 \caption{Change in the classical action for the backward sweep with zero initial action (i.e., $\Delta I=I_f-I_-=I_f$).
 Inset shows change in the action for small values of $\eps$ and a linear fit which gives $I_f/\eps=0.220631$.
 Theoretical prediction (\ref{backward}) is  $\frac{\ln  2}{\pi} \approx 0.220636$.
 }
\end{figure}
For a forward sweep, a classical phase point with $I_-=0$ (i.e.,
located initially at $n=0$) always remains there. Therefore, when
approximating an initial quantum state by an ensemble of classical
trajectories, one needs to carefully take into account quantum
fluctuations. For a ground state as an initial state, initial
actions of the classical ensemble $\sim 1/N$. For a backward
sweep, the situation is different. A phase point with $I_-=0$ is not
stationary and therefore in the semiclassical limit, one may
merely study its classical dynamics. For small sweeping rates,
dynamics remains adiabatic until the phase point approaches a
saddle point at $n=0$. Then, following \cite{ITorma}, close to a
saddle point the system can be mapped to a Painleve equation. To
this end, one introduces new variables $P= 2\sqrt{n} \cos \phi/2$,
\quad $Y=2 \sqrt{n} \sin \phi/2 $. After the
transformations and approximations described in \cite{ITorma}  one obtains the
Hamiltonian $ H= -s \frac{Y^2}{2} + \frac{P^2}{2} +
\frac{Y^4}{2},$ where $s$ is a new time variable and $P,Y$ are
canonically conjugated variables. One can see that the effective
Hamiltonian leads to the Painlev\'{e}-2 equation
\cite{Painleve,Its} \be \frac{d^2 Y}{ds^2}= sY - 2 Y^3.
\label{Penleve} \ee Asymptotics of the Painlev\'{e}-2 equation
were investigated by Its and Kapaev \cite{Its}. At $s \to -\infty$
the asymptotic solution to (\ref{Penleve}) is \cite{Its,Mitya} $$
Y(s)= \alpha (-s)^{-\frac{1}{4}} \sin \left( \frac{2}{3}(-s)^{3/2}
+ \frac{3}{4} \alpha^2 \ln(-s) + \varphi \right),$$ and in the
limit $s \to + \infty$ it is
 $$
Y(s) = \pm \sqrt{\frac{s}{2}}  \pm \rho(2s)^{-\frac{1}{4}} \cos
\left( \frac{2 \sqrt{2}}{3} s^{3/2} - \frac{3}{2} \rho^2 \ln(s) +
\theta \right),$$
 where $(\alpha, \varphi)$ and $(\rho,\theta)$ are
the "action-angle" variables characterizing the solutions in the
limits $s \to \pm \infty$. These variables are related by the
following relations \cite{Its,Mitya}:
\bea \rho^2 &=& \frac{1}{\pi} \ln \frac{1+|p|^2}{2 |\mbox{Im}(p)|} \nonumber\\
\theta &=& -\frac{\pi}{4} +\frac{7}{2} \rho^2 \ln 2 - \arg \Gamma(i \rho^2)-\arg(1+p^2), \label{relations}\\
 p&=& \left(\exp(\pi \alpha^2) \right)^{1/2} \times \nonumber\\
 &\times& \exp \left( i \frac{3}{2} \alpha^2 \ln 2 - \frac{\pi}{4}i - i \arg
 \Gamma\left(i \frac{\alpha^2}{2} \right)-i \varphi \right),\nonumber
 \eea
where $\arg(\cdot)$ denotes the argument of a complex number, and
$\Gamma(x)$ is the gamma function. As $s \to \pm \infty $, the
adiabatic invariant $I_p$ of the system (\ref{Penleve}) approaches
the quantities $I_p^-$ or $I_p^+$ which are defined (to the lowest
order terms) as \be I_p^-=\frac{\alpha^2}{2}, \quad I_p^+
=\frac{\rho^2}{2}. \label{relations2} \ee Relations
(\ref{relations}),(\ref{relations2}) allow to calculate the change in
the action both for a forward \cite{ITorma} and backward sweep.

It turns out that the backward sweep through the resonance has
very different behaviour from the forward one. Most importantly,
in the limit $I_- \to 0$ change in the action becomes
phase-independent. That is, at large $I_- \gg \eps$ predictions of
the Painlev\'{e} mapping correspond to the separatrix crossing
theory calculations. However, as $I_-$ decreases, the change in
the action approaches the phase independent value, which is
linearly proportional to $\eps$.

Indeed, for a classical trajectory with $I_-=0$ from
 (\ref{relations}),(\ref{relations2}) it follows
\cite{Mitya} \be I_p^- =\frac{1}{2\pi} \ln(1+|p|^2) \ee For the
backward sweep, $I_p^-$ corresponds to $I_f$, while $I_p^+$
corresponds to $I_-$. For the initial action $I_-=0$ we have
$|p|=0$. Taking into account all the transformations leading to
the Painlev\'{e} equation, we obtain: \be
 \Delta I =I_f= \frac{\ln 2}{ \pi} \eps \label{backward}
 \ee
 This is the main result of the present paper. Numerical
 calculations (see figure 2) reproduce the coefficient $\frac{\ln
 2}{\pi}$ with 5-digit accuracy. In Ref. \cite{Javanainen}, a linear power-law for deviation
  from adiabaticity was also found numerically,  however the accuracy of the analytical solution
  reported was lower than in our case.

To conclude, we found that the mapping of the many-particle
Landau-Zener model to the Painlev\'{e} equation \cite{ITorma}
enables to describe not only "forward" sweep through the
resonance, but also the reverse driving process. The latter
process arises, for example, when the model is used to describe
formation of diatomic molecules from Bose atoms, or dissociation
of diatomic molecules on Fermi atoms. Our results are relevant for
accurately describing the Feshbach resonance passage in ultracold
Fermi and Bose gases. Interferometric experiments with ultracold
quantum gases \cite{PNature1} is another possible area of
applications. Finally, we believe the method will have important
applications in the field of {\em dynamics of quantum phase
transitions} (QPT) \cite{QPTPolkovnikov,QPTZoller,PNature2}.

This work was supported by the Academy of Finland (Projects No.
213362, No. 217041, No. 217043, and No. 210953) and conducted as a
part of a EURYI scheme grant (see www.esf.org/euryi). A.P.I. was
partly supported by RFBR 06-01-00117.  A.P.I. is indebted to A.I.
Neishtadt for illuminating discussions.


\begin{thebibliography}{<30>}
\bibitem{footnote2}
Ref. \cite{Javanainen} considered system \{$i
\dot{\alpha}=\frac{\Omega}{\sqrt{2}} \alpha^* \beta, \quad i
\dot{\beta}=\frac{\Omega}{\sqrt{2}}\alpha^2+\delta \beta$ \}, with
$|\alpha|^2=1$ corresponding to the "all-atom" mode. Denoting
$n=|\alpha|^2$, $\phi= 2 \arg(\alpha)-\arg(\beta)$, after simple
rescalings we get the Hamiltonian (\ref{Polkovnikov}).
\end{thebibliography}
\end{document}